%% file: neuro.tex
\documentclass[10pt,conference]{IEEEtran}

\usepackage{latexsym}
\usepackage{amssymb}
\usepackage{amsmath}
\usepackage{mathrsfs}
\usepackage{amsfonts}
\usepackage{amsthm}
\usepackage{epsfig}
\usepackage{graphicx,psfrag,subfigure}

\input{definitions}

\def\be{\begin{eqnarray}}
\def\ee{\end{eqnarray}}
\def\ben{\begin{eqnarray*}}
\def\een{\end{eqnarray*}}

\title{
	From the Entropy to the Statistical Structure\\
	of Spike Trains
}

\author{
	\authorblockN{Yun Gao}
	\authorblockA{Brown University}
\and
	\authorblockN{Ioannis Kontoyiannis}
	\authorblockA{Athens University of Econ \& Business}
\and
	\authorblockN{Elie Bienenstock}
	\authorblockA{Brown University}
}

\begin{document}

\maketitle

\begin{abstract}
We use statistical estimates of the entropy rate of 
individual spike train data in order to make inferences 
about the underlying structure of the spike train itself. 
We first examine a number of different parametric and 
nonparametric estimators (some known and some new), 
including the ``plug-in'' method, several versions 
of Lempel-Ziv-based compression algorithms, 
a maximum likelihood estimator tailored to renewal processes, 
and the natural estimator derived from the Context-Tree Weighting 
method (CTW).
The theoretical properties of these estimators are examined,
several new theoretical results are developed, and all
estimators are systematically applied to various types 
of simulated data under different conditions.

Our main focus is on the performance of these entropy estimators 
on the (binary) spike trains of 28 neurons recorded simultaneously 
for a one-hour period from the primary motor and dorsal premotor
cortices of a monkey. We show how the entropy estimates
can be used to test for the existence of long-term structure 
in the data, and we construct a hypothesis test for whether 
the renewal process model is appropriate for these spike trains.
Further, by applying the CTW algorithm
we derive the maximum a posterior (MAP)
tree model of our empirical data, and
comment on the underlying structure it
reveals.
\end{abstract}

\section{Introduction}
%%%%%%%%%%%%%%%%%%%%%%%%%%%%%%%%%%%%%%%%%%%%%%%%%%%%%%%%%%%%%%%%%%%%%%
Information-theoretic methods have been widely 
used in neuroscience, in the broad effort to analyze 
and understand the fundamental information-processing
tasks performed by the brain.  In these studies,
the entropy has 
been adopted as a central measure for quantifying 
the amount of information transmitted between neurons.
One of the most basic
goals is to identify appropriate methods that can be used 
to estimate the entropy of spike trains 
recorded from live animals.

The most commonly used entropy-estimation 
technique is probably the so-called ``plug-in'' 
(or maximum-likelihood) 
estimator and its
various modifications.  This method consists of essentially
calculating the empirical frequencies of all words
of a fixed length in the data, and then 
estimating the entropy rate of the data by
calculating the entropy of this empirical distribution;
see, e.g., 
\cite{Strong:98}\cite{paninski:03}\cite{warland:97}\cite{Pamela:00}%
\cite{Stevens:96}\cite{Nemenman:04}\cite{deRetal}.
For computational reasons, 
the plug-in estimator cannot go beyond word-lengths 
of about 10 or 20, and hence it does not take into
account the potential longer time dependence in the signal. 

Here we examine the performance of various 
entropy estimators, including some based on
the Lempel-Ziv (LZ) data compression algorithm 
\cite{ziv-lempel:1},
% \cite{Ziv:78}, 
and
some based on the Context-Tree Weighting (CTW)
method for data compression.
We employed four different LZ-based methods;
of those, two
% \cite{Kontoyiannis:96}
\cite{Kontoyiannis:98},
have been widely 
and very successfully used in many applications
(e.g.,
\cite{schu-grass:96}\cite{Kontoyiannis:98}),
and the other two are new estimators with some 
novel and more desirable statistical properties.
The CTW-based estimator we used
is based on the results in
\cite{willems:95}\cite{willems:98}
and it has also been considered in 
\cite{Kennel:02}\cite{London:02}.

We demonstrate that the LZ- and CTW-based 
estimators naturally incorporate dependencies 
in the data at much longer time scales, and that
they are consistent (in the statistical sense) 
for a wide class of data types generated from 
distributions that may possess arbitrarily long 
memory.

To compare the performance of various methods, we applied
these entropy estimators on simulated data
generated from a variety of different processes,
with varying degrees of dependence. 
% in all cases
% the true entropy rate can be precisely calculated 
% or accurately computed.
We study the convergence rate of the bias and variance
of each estimator, and their relative performance
in connection with the length of the memory present 
in the data.
\comment{
Our analysis shows that, whereas for 
short-memory processes the plug-in is as good
as any other method, for processes with longer 
memory the plug-in is much worse than
both the LZ estimators and the CTW, because of
undersampling problem.  In fact,
the CTW estimator is uniformly better than the other estimators,
for both short and relatively long memory process.  
Its fast convergence rate outperforms the LZ-based estimators, and
its ability to allow for longer memory makes it more accurate
than the plug-in.
}

Finally, we applied these methods to neural data,
recorded from
two multi-electrode arrays implanted on a monkey's 
primary motor cortex (MI) and dorsal premotor cortex (PMd).
The arrays
simultaneously recorded neural activity from 
28 different neurons.  A Plexon acquisition system
was used to collect neural signal, and the units were 
spike-sorted using Plexon's Offline Sorter.
The monkey was not engaged in any task when the
data were collected, and the size of the data 
is approximately an hour. 
A detailed description of recording techniques 
is given in \cite{Maynard:99}. 

Our {\sc\bf main conclusions} can be summarized as follows:

$\bullet\;$~The CTW was consistently the most reliable and 
accurate estimator.

$\bullet\;$~The results of the CTW compared with those of the
plug-in method very strongly suggest that there are 
significant longer-term dependencies in the data.

$\bullet\;$~One of the most significant features of our 
results is the observation that from the CTW algorithm 
we can also obtain an explicit statistical model for the 
data, the so-called 
{\em maximum a posteriori probability tree} model
\cite{maxtree1}. From the resulting tree structures 
we deduce several interesting aspects
of spike train patterns. In particular, we find that
the primary statistical feature of a spike train 
captured by the CTW estimator is its empirical 
inter-symbol interval (ISI)
distribution.

$\bullet\;$~Among all the estimators we considered,
the CTW with large depth $D$ is the only method able 
to capture the longer term statistical structure 
of renewal data with ISI distribution that is close 
to that of real neurons.

$\bullet\;$~The spike train data we examined can,
to a statistically significant degree of accuracy,
be modelled as renewal processes with independent
ISIs. Specifically, in the entropy estimation task,
among the various sources of bias,
% (such as CTW's inherent upward bias, and the negative 
% bias due to undersampling), 
the bias incurred 
by treating the spike train as a renewal process 
is negligible for the neurons considered.

$\bullet\;$~In our entropy estimation experiments 
on spike trains we generally observed that, as the 
tree depth of the CTW increases, the corresponding 
entropy estimates decrease. This decrease is 
significantly larger than would be expected from
purely random fluctuations, if there was actually 
{\em no} long-term structure in the data.

$\bullet\;$~The percentage 
of drop is correlated with the variability 
of the spike count in the data (as quantified
by the Fano factor). This conclusion is 
rigorously justified
with a $t$-test with a $p$-value 
of $\approx 5\times 10^{-4}$. Perhaps most
interestingly, since the test was done on
data in 100ms windows, it implies 
that the correlation is {\em not} simply due to
refractoriness or any other structure that appears
on a fine time scale.

\section{Experimental Results and Findings}
%%%%%%%%%%%%%%%%%%%%%%%%%%%%%%%%%%%%%%%%%%%%%%%%%%%%%%%%%%%%%%%%%%%%%%
\label{sec:data}

\begin{figure}[ht]
\psfig{figure=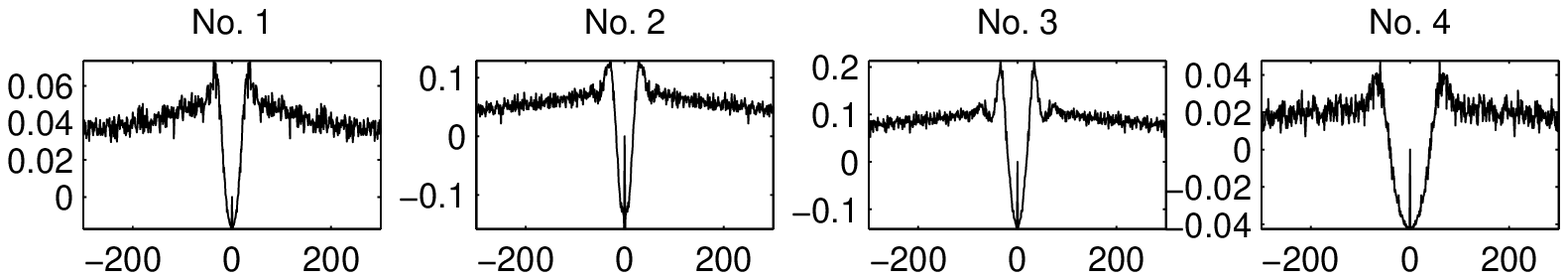,width=3.3in}
% \centerline{
% \psfig{figure=auto2.eps,width=3.4in}
% }
\psfig{figure=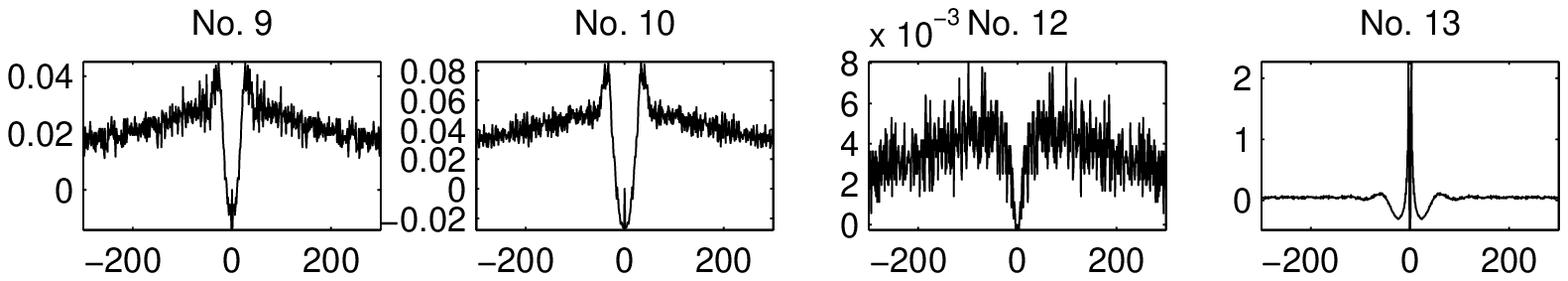,width=3.3in}
% \centerline{
% \psfig{figure=auto4.eps,width=3.4in}
% }
\psfig{figure=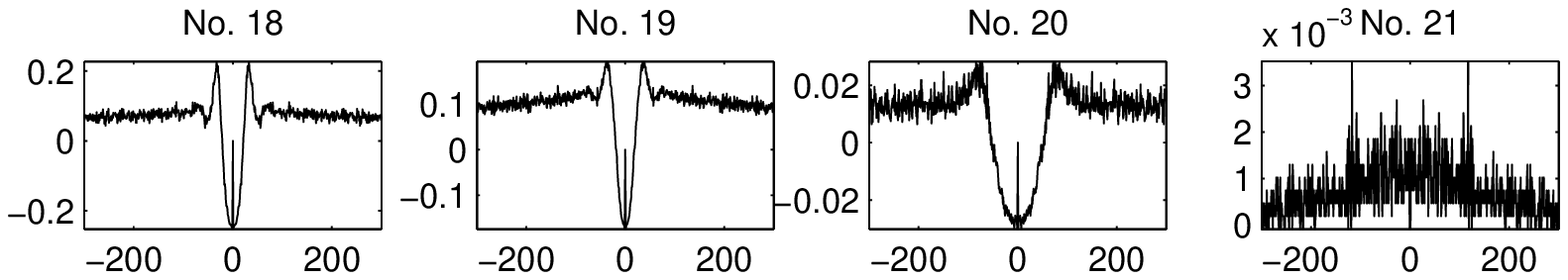,width=3.3in}
\psfig{figure=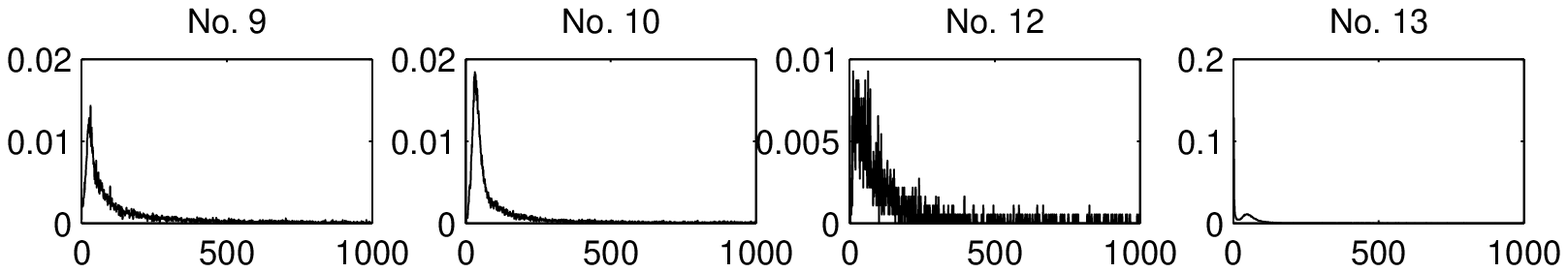,width=3.4in}
% \centerline{
% \psfig{figure=auto6.eps,width=3.4in}
% }
% \centerline{
% \psfig{figure=auto7.eps,width=3.4in}
% }
\vspace*{-0.1in}
\caption{
The first three rows show the autocorrelograms of 12 out of the 28 
spike trains from 28 different neurons.
Lag varies from -300ms to 300ms.
The last row shows the empirical ISI distributions of 4 spike trains
from 4 of the 28 neurons. ISI values vary from zero to
1000ms.
}
\label{fg:autocorr}
\end{figure}

Our neuronal data come from binned spike trains
recorded simultaneously from 28 neurons, with bin 
size equal to 1ms. The total length of each spike 
train is $N=3,606,073$ms, which is a little over 
an hour.
Five out of the 28 neurons have
average firing rates lower than 1Hz, 
16 are between 1Hz and 10Hz, 
6 are between 11Hz and 20Hz, 
and one has a firing rate above 20Hz. 
Figure~\ref{fg:autocorr} shows the 
autocorrelograms for 12 of the 28 neurons
and 
the empirical inter-symbol interval (ISI) 
distributions for 4 of the 28 neurons.
These plots show that there is great richness and
variability
in the statistical behavior of different neurons.

% \begin{figure}[ht]
% \psfig{figure=isi1.eps,width=3.4in}
% \centerline{
% \psfig{figure=isi2.eps,width=3.4in}
% }
% \centerline{
% \psfig{figure=isi4.eps,width=3.4in}
% }
% \psfig{figure=isi5.eps,width=3.4in}
% \centerline{
% \psfig{figure=isi6.eps,width=3.4in}
% }
% \centerline{
% \psfig{figure=isi7.eps,width=3.4in}
% }
% \vspace*{-0.1in}
% \caption{
% }
% \label{fg:empISI}
% \end{figure}

\subsection{Entropy Estimates}

\begin{figure}[ht]
\psfig{figure=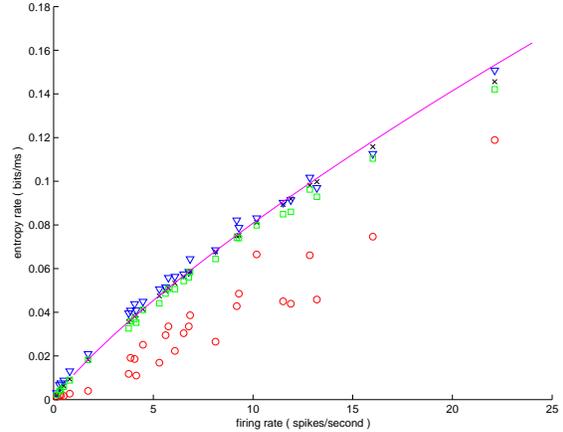,width=2.9in}
\vspace*{-0.1in}
\caption{
Entropy estimates obtained by several methods, shown 
in bits-per-ms and plotted against the mean firing rate
of each neuron.
The purple curve is the entropy rate of an i.i.d.\
process with the corresponding firing rate.
The results of the plug-in (with word-length $20$)
are shown as black x's; 
the results of the two LZ-based estimators as
red circles and blue triangles;
and the results of the CTW method as green squares.
% CTW with $D=\infty$.
}
\label{fg:neuraldata}
\end{figure}

Figure~\ref{fg:neuraldata} shows the results
of estimating the 
entropy rate of these spike 
trains, using the plug-in method, two LZ estimators,
and the CTW algorithm.

Note that all the estimates generally increase as the
firing rate goes up, and that the i.i.d.\ curve
corresponds exactly to the value to which the plug-in 
estimator with word length $1$ms would converge with
infinitely long data. The plug-in estimates with word length 
$20$ms are slightly below the i.i.d.\ curve,
and the CTW estimates tend to be slightly lower than 
those of the plug-in. It is important to observe
that, although the bias of the plug-in is negative 
and the bias of the CTW is positive, we consistently
find that the CTW estimates are {\em smaller} than 
those of the plug-in. This strongly suggests that 
the CTW {\em does indeed find significant longer-term 
dependencies in the data}. 

For the two LZ estimators, we observe that one
gives results that are systematically higher 
than those of the plug-in, and the other
is systematically much lower.
The main limitation of the plug-in is that it can only
use words of length up to 20ms, and even for word lengths 
around 20ms the undersampling problem makes these estimates
unstable. 
Moreover, this method completely misses the effects of longer 
term dependence. Several {\em ad hoc}
remedies for this drawback have been
proposed in the literature; see, e.g.,
\cite{Strong:98}.
% which begins with the plug-in for relatively short word-lengths
% and extrapolates in an {\em ad hoc} fashion to ``infinite'' 
% word-lengths, which the authors suggest should correspond 
% to the ``true'' entropy rate of the underlying process.
% On the other hand, the CTW method and the LZ-based estimators
% take longer term dependence into account in a rigorously
% justifiable way.

The main drawback of the LZ estimators is 
the slow rate of convergence of their bias, which 
is relatively high and hard to evaluate analytically.
% On the other hand, their variance is relatively small.

As we found in extensive simulation studies, the bias of 
the CTW estimator converges much faster than the biases
of the LZ estimators, while keeping the advantage
of dealing with long-range dependence.  Moreover, 
from the CTW we can obtain an explicit statistical 
model for the data, the ``maximum a posteriori probability'' 
(MAP) tree described in \cite{maxtree1}.
% In our study we looked extensively at the 
% MAP trees resulting from spike-train data.
The importance of these
models comes from the fact that, in the 
information-theoretic context, they can be 
operationally interpreted as the ``best'' 
tree models for the data at hand.

\subsection{MAP Tree Models for Spike Trains}
\label{subsec:treestruct}
%%%%%%%%%%%%%%%%%%%%%%%%%%%%%%%%%%%%%%%%%%%%%%%%%%%%%%%%%%

\begin{figure}[ht]
\centerline{
\psfig{figure=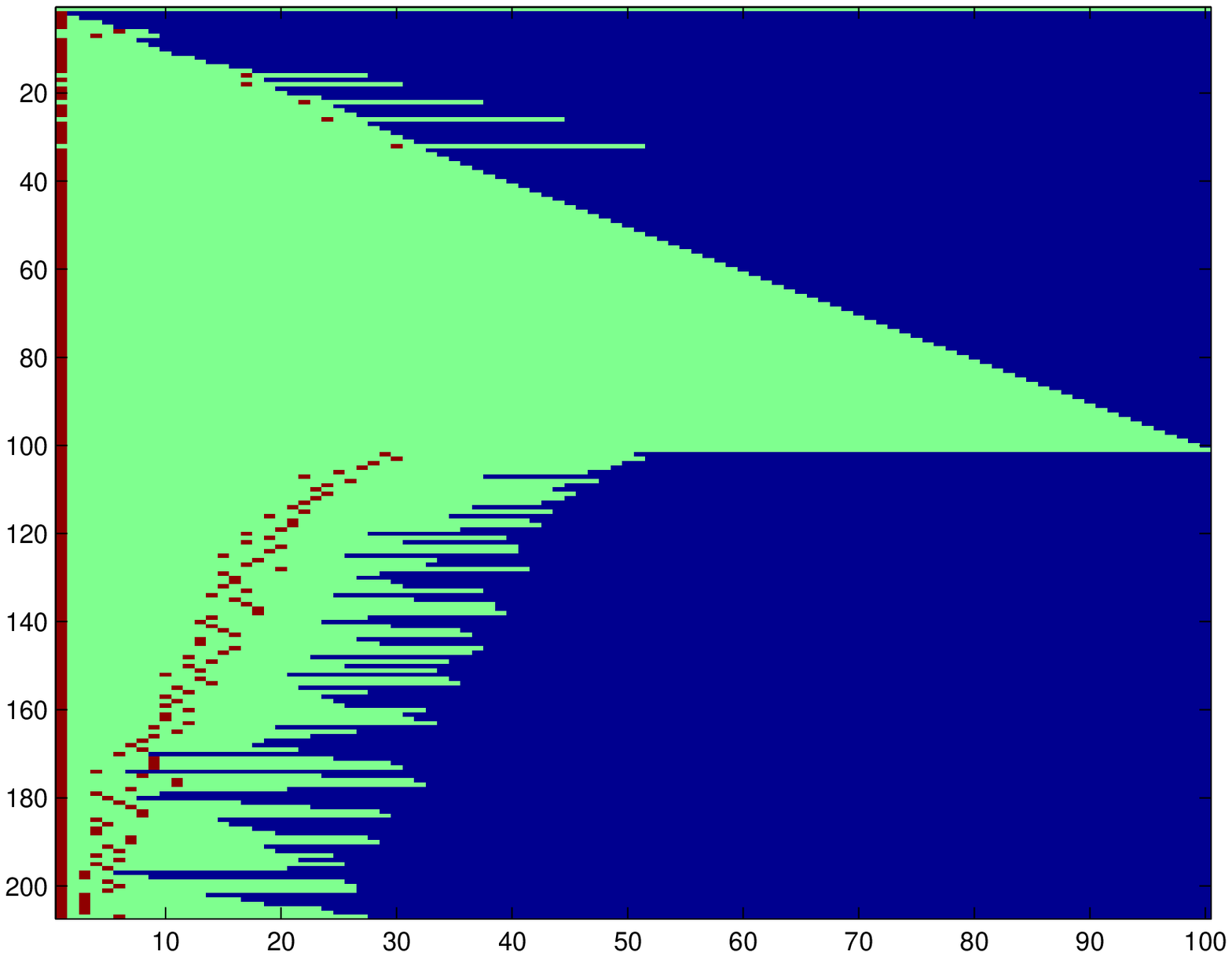,width=2in}
}
\centerline{
\psfig{figure=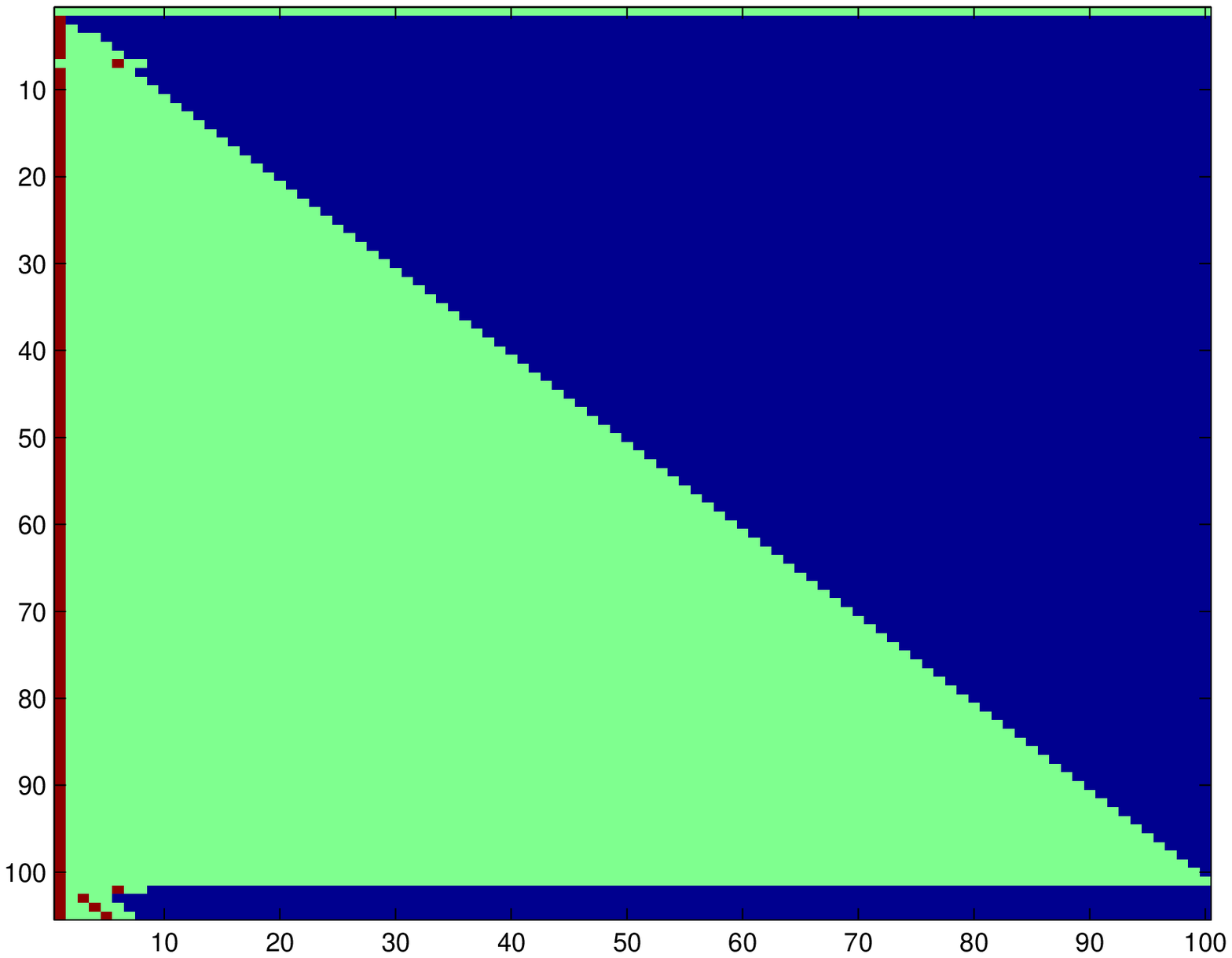,width=2in}
}
\vspace*{-0.1in}
\caption{Suffix sets of MAP trees derived from the
spike trains of cells 1 and 4, whose mean rates are
4.15Hz and 6.52Hz, respectively.  Suffixes are sorted
in descending order of frequency. The green areas are zeros 
in the suffixes, red dots are 1's, and the blue areas
mark the end of each suffix.
}
\label{fg:trees}
\end{figure}

We computed the MAP tree models 
\cite{maxtree1}
derived from 
spike train data 
using the CTW algorithm
with depth $D=100$.
Figure~\ref{fg:trees} shows the suffix sets of two
cells' MAP trees, sorted in descending order of suffix 
frequency.  The most frequent suffix is always
the all-zero suffix, generally followed by suffixes of the
form ``$1000000\cdots 0$.'' Similarly to the
results of Figure~\ref{fg:autocorr},
we find a lot of variability between neurons.

Since suffixes of the form ``$100\cdots 0$'' 
are the most common non-zero suffixes produced by
the CTW, 
% in Table~\ref{tab:isisuffix} we show 
% the percentage of such suffixes among all the 
% non-zero suffixes in the MAP trees we obtained 
% for each of the 28 neurons; 
we note that in 22 out of the
28 neurons
the percentage of such suffixes among all the 
non-zero suffixes in the MAP trees we obtained 
exceeds 75\%.
Since the frequency of each such suffix is exactly 
the same as the frequency of an inter-spike interval 
with the same length, we interpret the high frequency
of ``$100\cdots 0$'' suffixes as an indication that
the primary statistical feature of a spike train 
captured by the CTW estimator is its empirical 
inter-symbol interval (ISI)
distribution.

% \begin{table}
  % \begin{center} 
    % \begin{tabular}{|c|ccccccc|}
      % \hline
 % cell & \multicolumn{7}{|c|}{Percentage} \\
% \hline
% 1-7 & 88.73 & 75.62 & 51.06 & 98.73 & 100.00 & 99.37 & 55.99\\
% 8-14 & 78.81 & 97.65 & 89.02 & - & 98.69 & 47.30 & 97.29\\
% 15-21 & 98.77 & 0 & 77.00 & 57.03 & 45.02 & 98.20 & 100.00 \\
% 22-28 & 98.78 & 98.88 & 96.06 & 91.02 & 97.13 & 100.00 & 98.86\\
% 29 & 100.00 & & & & & & \\
% \hline
    % \end{tabular}
    % \vspace*{-0.1in}
% \caption{Percentage of suffixes
	% of the form ``$1000000\cdots 0$'' among 
	% all non-zero suffixes.
	% The MAP tree of cell 16 consists of only
	% the all-zero suffix.}
    % \label{tab:isisuffix}
  % \end{center}
% \end{table}

\begin{figure}[ht]
\psfig{figure=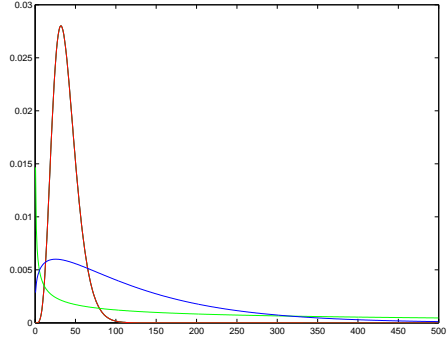,width=2.3in}
\vspace*{-0.1in}
\caption{The three components of the mixture of 
Gamma's fitted to cell 1. 
% Only the first 500ms are shown.
}
\label{fg:mixgamma2}
\end{figure}

This observation motivates us to look at renewal process 
models in more detail. Next we examine the performance of
different entropy estimators on simulated data from 
renewal processes whose ISI distribution is close to 
that of real spike trains. For that, we first
estimate the ISI distribution of our spike trains. 
Using the empirical ISI distribution
is problematic since such estimates are typically
undersampled and hence unstable.
Instead, we developed and used a simple iterative
method to fit a mixture of three Gamma's to the
empirical estimate. The idea is based on the
celebrated EM algorithm \cite{EM}.
% Figure~\ref{fg:mixgamma} shows the fitted densities 
% for three cells whose spike trains appear to have 
% rather different statistics -- the same 
% three cells whose suffix sets were shown in 
% Figure~\ref{fg:trees}. 
The components of the
mixtures can be though of as capturing different aspects of 
the ISI distribution of real neurons. For cell
no.\ 1, for example, Figure~\ref{fg:mixgamma2} shows
the shapes of these three components. One of them
is highly peaked around 40ms, probably due to refractoriness;
another component appears to 
take into account longer-range dependence,
and the last component adjusts some details 
of the shape.

% \begin{figure}[ht]
% \psfig{figure=mixgamma.eps,width=3.4in}
% \vspace*{-0.1in}
% \caption{Fitted mixture of three Gamma's  to the 
% empirical ISI distributions of three cells.  
% The $x$-axis is in ms.  The blue curve is the
% empirical density, the green curve is an initial 
% fit of the mixtures, and the red curve is the fit 
% after ten interations.
% }
% \label{fg:mixgamma}
% \end{figure}

\begin{figure}[ht]
\psfig{figure=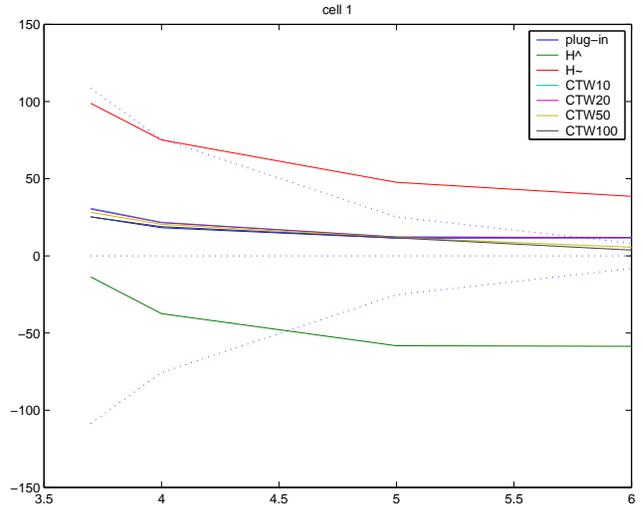,width=3.3in}
\vspace*{-0.1in}
\caption{Bias (as a percentage of true entropy rate) of various
entropy estimators applied to simulated data from a renewal 
process with ISI distribution given by a mixture
of Gammas fitted to the spike train of neuron no.\ 1. 
The data lengths are $N=5000$, $10^4$, $10^5$ and $10^6$,
and the $x$-axis is $\log_{10} N$.
The upper and lower dotted lines represent two standard 
deviations away from the true values, where the standard
deviation is obtained as the average of the standard errors 
of all the methods. The plug-in has word-length $20$ms,
% the two LZ estimators are 
% $\hat{H}_{n}$ and $\tilde{H}_{n}$, 
and the CTW estimator is used with four different tree 
depths, $D=10,20,50,100.$
The true entropy rate of this ``cell'' is 0.0347 bits/ms.
}
\label{fg:biasvarrenewal1}
\end{figure}

We then run various estimators on simulated data 
from renewal processes with ISI distributions given 
by the three-Gamma mixtures obtained from the spike
train data. 
Figure~\ref{fg:biasvarrenewal1}
% \ref{fg:biasvarrenewal2} 
% and~\ref{fg:biasvarrenewal3} 
shows the bias of various estimators 
as a percentage of the true entropy rate, 
for simulated data with ISI distributions 
given by the Gamma-mixtures obtained from 
cell 1. The data lengths
are $N=5000$, $10^4$, $10^5$ and $10^6$.
% Since the standard errors for several of the
% methods are very close, displaying them as 
% error bars around each estimate would obscure the plot.
% Instead, we calculated the standard error of each method,
% took their average, and plotted upper and lower bounds
% corresponding to two standard deviations of the 
% average standard error.

% \begin{figure}[ht]
% \centerline{
% \psfig{figure=cell4_report3.eps,width=3.4in}
% }
% \caption{Results as in Figure~\ref{fg:biasvarrenewal1},
% with ISI distribution fitted to the spike train of
% neuron no.\ 4. The true entropy rate is
% 0.0540 bits/ms.}
% \label{fg:biasvarrenewal2}
% \end{figure}

% \begin{figure}[ht]
% \psfig{figure=cell18_report3.eps,width=3.4in}
% \caption{Results as in Figure~\ref{fg:biasvarrenewal1},
% with ISI distribution fitted to the spike train of
% neuron no.\ 18. The true entropy rate is
% 0.1112 bits/ms.}
% \label{fg:biasvarrenewal3}
% \end{figure}

As we can see from the plot, the LZ estimators converge very
slowly, while the plug-in and CTW estimates are much more accurate.
The estimates obtained by the plug-in with word-length
$20$ms and by the CTW with depths $D$=10, 20 are very similar, 
as we would expect. But for data sizes $N=10^5$ or greater, the 
CTW with longer depth gives significantly more accurate results
that outperform all other methods. Similar comments apply to the
results for most other cells, and for some cells
the difference is even greater in that the CTW
with depth $D=20$ already significantly outperforms the plug-in
with the same word length.
The standard error of the plug-in and the CTW estimates 
is larger than their bias for small data lengths,
but at $N=10^6$ the reverse happens.

In short, we find that that the CTW with large depth $D$ is the
only method able to capture the longer term statistical structure 
of renewal data with ISI distribution that is close to that
of real neurons. For cell no.\ 1, for example, the largest ISI 
value in the data is 15.7 seconds.
Therefore, there is significant
structure at a scale that is {\em much} greater that the
plug-in window of 20ms, and the CTW apparently can take advantage
of this structure to improve performance.

Finally, we remark on how the choice of the depth $D$ affects the
CTW estimator; a more detailed discussion is given in 
Section~\ref{subsec:fano} below.
For smaller data sizes $N$, the results of the
CTW with $D=10$ are very close to those with $D=100$, 
but for $N=10^5$ and $N=10^6$ the difference becomes 
quite significant.
This is likely due to the fact that, for small $N$,
there are not enough long samples to represent the
long memory of the renewal processes, 
and the estimation bias is dominated 
by the undersampling bias. For large $N$, on the other hand,
undersampling problems become more minor, and the difference
produced by the longer-term dependence captured 
by larger depths becomes more pronounced.

\subsection{Testing the Non-renewal Structure of Spike Trains}
%%%%%%%%%%%%%%%%%%%%%%%%%%%%%%%%%%%%%%%%%%%%%%%%%%%%%%%%%%%%%%%%%%%%%%%%
\label{subsec:nonrenewal}

The above results, both on simulated data and on real 
neural data, strongly indicate that the CTW with large tree 
depth $D$ is the best candidate for accurately estimating 
the entropy rate of binned spike trains.
They also suggest that the main statistical pattern
captured in the CTW's estimation process is the 
renewal structure that seems to be inherently present 
in our data. It is then natural to ask how accurately 
a real spike train can be modelled as renewal process,
or, equivalently, how much is ``lost'' if we assume that
the data are generated by a renewal process.

Recall that, if a spike train 
has {\em dependent} ISIs, then its entropy rate will be 
{\em lower} than that of a renewal process with the same 
ISI distribution. Therefore we can specifically ask the
following question: If we took the data corresponding 
to a real spike train and ignored any potential dependence 
in the ISIs, would the estimated entropy differ significantly 
from the corresponding estimates using the original data?

\begin{figure}[ht]
\psfig{figure=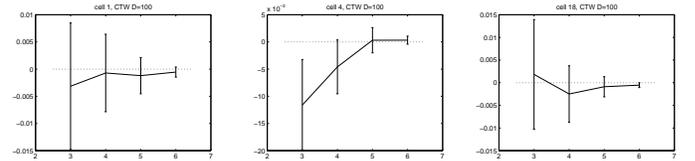,
width=3.5in}
\vspace*{-0.1in}
\caption{Difference of entropy estimates on spike trains and on
simulated data from a renewal process with the same ISI distribution.
The segment lengths are $N=10^3$,$10^4$, $10^5$, and $10^6$. The 
$x$-axis is $\log_{10}{N}$ and the $y$-axis is the difference 
between the estimates in bits/ms. For each $N$, 
a hundred randomly chosen segments from real data 
(and the same number of realizations of simulated data)
are used to compute the estimates, using the CTW 
algorithm with $D=100$.  The difference is the average of 
the 100 estimates based on real data minus the corresponding 
average on simulated data.  The error bar is two standard 
deviations of the estimated difference.
}
\label{fg:nonrenewal}
\end{figure}

To answer the above questions we performed the following 
experiment. For a given neuron, we estimated the ISI 
distribution by a three-Gamma mixture as described
earlier and generated a renewal process with that ISI 
distribution. Then we randomly selected long segments
of fixed length from the simulated data and estimated
the entropy using the CTW algorithm with depth $D=100$; 
the same was done with the real spike train.
Finally, we compared the average value of the
estimates from the simulated data set to 
that of the spike train estimates.
We used 100 segments from each data set,
of lengths $N=1000$, $10^4$, $10^5$ and $10^6$. 
Figure~\ref{fg:nonrenewal} shows the 
difference between the two (averaged) estimates,
that is, the estimated entropy of the real spike
train minus the estimated entropy of the 
corresponding renewal process. We also plot 
error bars corresponding to two standard 
deviations of this difference.
% , with the standard
% deviation calculated as,
% $$\bar{\sigma}_{\rm difference}=
% \sqrt{\frac{1}{n}\,\bar{\sigma}_{\rm spike-train}^2
	% +\frac{1}{n}\,\bar{\sigma}_{\rm renewal}^2}\;,$$
% where $n=100$. 
The same experiment was performed on data
from three different cells.

From the plots we can see that at smaller $N$ the error bars 
are very wide and the estimated difference well within two
standard deviations away from zero. Even at $N=10^6$ where
the difference becomes more apparent, it remains very close
to the two-standard-deviation bound. Hence the results
suggest that either there is no significant difference,
or, if it exists, it is rather negligible. In other words,
among the various sources of bias in the entropy estimation
process (such as CTW's inherent upward bias, the negative 
bias due to undersampling, and so on), the bias incurred 
by treating the spike train as a renewal process is negligible 
for the neurons we examined.

\subsection{Memory Length in Spike Trains}
%%%%%%%%%%%%%%%%%%%%%%%%%%%%%%%%%%%%%%%%%%%%%%%%%%%%%%%%%%%%%%%%%%%%%%%%%
\label{subsec:fano}

A natural measure of the amount of dependence 
or long-term memory in a process $\{X_n\}$,
is the rate at which the conditional entropy 
$H(X_n|X_{n-1},\ldots,X_1)$ converges to 
the entropy rate $H$.
% similarly for the rate at which the average per-sample
% entropy $\frac{1}{n}H(X_1,\ldots,X_n)$ converges.
We can relate this rate of decay to the tree depth 
of the CTW algorithm in the following way. If the data
are generated from a tree source with depth $D$
(or any source with memory-length that 
does not exceed $D$), then the CTW estimator will 
converge to the true entropy rate, which, in this case, 
is equal to the conditional entropy,
$H(X_{D+1}|X_{D},\ldots,X_1).$
If, on the other hand, the data comes from a process
with longer memory, the estimates will still converge
to 
$H(X_{D+1}|X_{D},\ldots,X_1),$
but this will
be strictly larger than the actual entropy rate.
Therefore, in principle, we could perform the following
experiment: Given unlimited
data and computational resources, we could estimate the
entropy using the CTW with different tree depths $D$.
As $D$ increases the estimates will decrease, 
up to the point where we reach the true memory length
of the process, after which the estimates will remain
constant. Of course in practice we are limited
by the length of the data available (which adds 
bias and variability), 
and also by the amount of computation we can perform 
(which restricts the range of 
$D$'s we may consider). Nevertheless, for the case
of the spike train data at hand, some conclusions
can be drawn with reasonable confidence.

\begin{figure}
\psfig{figure=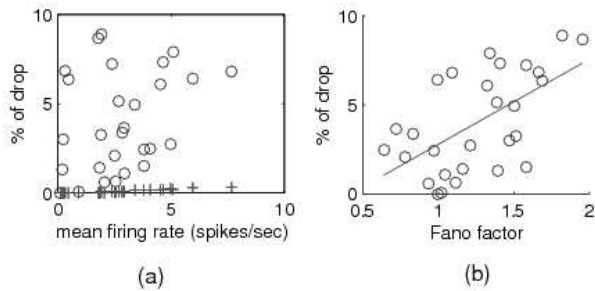,width=3.2in}
% \vspace*{-0.1in}
\caption{(a)~Percentage of drop in the CTW entropy
estimates from $D=1$ to $D=100$, plotted against
each neuron's firing rate. Blue circles denote the results
on the 28 neural spike trains. Red crosses are corresponding
results on simulated i.i.d.\ data with the same mean 
firing rates as real neurons.
(b)~Scatter plot of the Fano factors of real neurons
(with a 100ms window) plotted against the percentage of 
drop in their entropy estimates.
% (c) Fano factor vs. percentage of drop of generated spike trains from
% inhomogeneous Poisson with slowly varying rate.
}
\label{fg:ctwdrop}
\end{figure}

In our experiments on spike trains
we generally observed that, as the tree depth $D$ of the CTW
increases, the corresponding entropy estimates decrease;
see Figure~\ref{fg:biasvarrenewal1}.
% ,~\ref{fg:biasvarrenewal2}
% and~\ref{fg:biasvarrenewal3}.
However, the percentage of drop 
from $D=1$ to $D=100$ varies greatly 
from neuron to neuron, ranging from 
0\% to 8.89\%, with a mean of 4.01\% and 
standard deviation 2.74\%.
These percentages are shown as blue circles in
Figure~\ref{fg:ctwdrop}(a).
Since it is not {\em a priori} clear
whether the drop in the entropy estimates 
is really due to the presence of longer
term structure or simply an artifact 
of the bias and variability of the estimation 
process,
in order to get some measure for comparison
we performed the same experiment on a 
memoryless process:
We generated i.i.d.\ data with the same
length and mean firing rate as each of the neurons, 
and computed the percentage of drop in 
their entropy estimates from $D=1$ to $D=100$; 
the corresponding results are plotted
in Figure~\ref{fg:ctwdrop}(a) as red crosses.

From the plots we clearly see that the drop
in the entropy estimates on i.i.d.\ data
is significantly smaller and
much more uniform across neurons, 
compared to the corresponding results
on the spike train data. We next
investigate the potential reasons
for this drop, and ask whether these
results indicate that some neurons have 
more long-term structure and longer time 
dependency than others. An important quantity
for these considerations -- often used to 
quantify the variability of a spike train  --
is the Fano factor. This is defined as the
ratio of the variance of the number of spikes 
counted in a specified time window, to
the average spike count in such a window;
see, e.g., \cite{spikesbook}.  
Figure~\ref{fg:ctwdrop}(b) 
shows the scatter plot of the Fano factors 
of real neurons (computed with a 100ms bin size) 
against the percentage of drop in their
entropy estimates. At first glance, at least,
they appear to be positively correlated. 
To quantify the significance of this 
observation we use a $t$-test:
% \cite{statextbook}: 
The null hypothesis is that the 
correlation coefficient between the Fano factors
and the percentage drop is zero, and the
test statistic is 
$t=\frac{r\sqrt{n-2}}{\sqrt{1-r^2}} \sim t_{n-2},$
where $r$ is the sample correlation and $n$ is the number
of samples; here $r=0.5843$, $n=28$. The test gives $p$-value
of $5.4802\times 10^{-4}$, which means that we can confidently 
reject the null, or, alternatively, that the Fano
factors and the entropy drops are positively correlated.

Recall \cite[pp.52-53]{spikesbook},
that
the Fano factor of i.i.d.\ data
(i.e., data from a renewal process 
with ISIs that are geometrically 
distributed, often referred to as
``Poisson data'') is exactly equal to 1, 
whereas for most of the neurons with large percentage
of drops we find Fano factors greater than 1. This is
further indication that in neurons with larger percentages
of drop in the entropy estimates we see greater departure
from i.i.d.\ firing patterns. We should also note that,
since the Fano factor with a 100ms bin is completely blind
to anything that happens in shorter time scales,
this is {\em not} a departure from i.i.d.\ firing
in terms of the fine time structure. Among potential 
explanations we briefly 
mention the renewal structure of the data with perhaps 
long tails in the ISI distribution, and also 
the possibility of a ``slow''
modulation of the firing rate, creating longer memory
in the data, although the latter explanation cannot
explain the heavy-tailed nature of the ISIs
observed.
% To test this theory, we simulate spike trains
% from inhomogeneous Poisson with rate function:
% $r(t)=r_0+A cos(2\pi/Tt)$, where $r_0$ is the base rate (here taken
% to be the mean rate of real neurons), amplitude $A$ is chosen
% so that the rate has some degree of variability, 
% period $T$ here taken to be 10 times the bin 
% size,i.e. 1000ms.
% Figure \ref{fg:ctwdrop}(c) shows the scatter plot
% of Fano factor against percentage of drop.  
% As we can see, all the $FF$s are greater than 1, and significantly
% related with drops, $r=0.9323$, $p$-value=2.7123e-13.
% Hence we can conclude that a slowly varying rate is a very likely
% cause of the drop
% of the CTW estimate as longer history is taking into account.

\subsubsection*{Acknowledgments}
{\small
Y.G.\ was supported by the Burroughs Welcome fund.  
I.K.\ was supported by a Sloan Foundation Research 
Fellowship and by NSF grant \#0073378-CCR.
E.B.\ was supported by NSF-ITR 
Grant \#0113679 and NINDS Contract N01-NS-9-2322.
We thank Nicho Hatsopoulos for providing 
the neural data set.
}

{\small

%%%%%%%%%%%%%% end of insertion %%%%%%%%
\bibliographystyle{plain}

\def\mybibitem{\vspace{-0.4in}\bibitem}

}

\end{document}

%% file: definitions.tex
\newcommand{\comment}[1]{}

\def\0{{\bf 0}}

\def\a1{\mbox{\bf a}_1}
\def\a2{\mbox{\bf a}_2}
\def\a3{\mbox{\bf a}_3}
\def\a4{\mbox{\bf a}_4}

%% file: neuro.bbl
\begin{thebibliography}{10} \setlength{\itemsep}{-0.3ex}
% \small

\vspace{-0.1in}

\bibitem{deRetal}
R.R.\/ de Ruyter van Steveninck, et. al.
\newblock Reproducibility and variability in neural spike trains.
\newblock {\em Science}, 275:1805--1808, 1997. 

\bibitem{EM}
A.P.\/ Dempster, N.M.\/ Laird and D.B.\/ Rubin.
\newblock Maximum likelihood from incomplete data via the EM algorithm 
(with discussion). 
\newblock {\em Journal of the Royal Stat Society B}, 39:1--38, 1977.

\bibitem{Kennel:02}
M.\/ Kennel and A.\/ Mees.
\newblock Context-tree modeling of observed symbolic dynamics. 
\newblock {\em Phys. Rev. E}, 66:056209, 2002.

% \bibitem{Kontoyiannis:96}
% I.\/ Kontoyiannis.
% \newblock The complexity and entropy of literary styles.
% \newblock {\em NSF Technical Report No. 97}, Department of Statistics, 
% Stanford University, June 1996.

\bibitem{Kontoyiannis:98}
I.\/ Kontoyiannis, P.H.\/ Algoet, Yu.M.\/ Suhov and A.J.\/ Wyner.
\newblock Nonparametric entropy estimation for stationary processes and random fields, with applications to English text.
\newblock {\em IEEE Trans. Inform. Theory}, 44:1319--1327, 1998.

\bibitem{London:02}
M.\/ London.
\newblock The information efficacy of a synapse.
\newblock {\em Nature Neurosci.}, 5(4):332--340, 2002.

\bibitem{Maynard:99}
E.\/ Maynard, N.\/ Hatsopoulos, C.\/ Ojakangas, B.\/ Acuna,
  J.\/ Sanes, R.\/ Normann and J.\/ Donoghue.
\newblock Neuronal interaction improve cortical population coding of movement direction.
\newblock {\em J. of Neuroscience}, 19(18):8083--8093, 1999.

\bibitem{Nemenman:04}
I.\/ Nemenman, W.\/ Bialek and R.\/ de Ruyter van Steveninck.
\newblock Entropy and information in neural spike trains: progress on
the sampling problem.
\newblock {\em Physical Review E}, 056111, 2004.

\bibitem{paninski:03}
L.\/ Paninski.
\newblock Estimation of entropy and mutual information. 
\newblock {\em Neural Comp.}, 15:1191--1253, 2003.

\bibitem{Pamela:00}
P.\/ Reinagel.
\newblock Information theory in the brain.
\newblock {\em Current Biology}, 10(15):R542--R544, 2000.

\bibitem{spikesbook}
F.\/ Rieke, D.\/ Warland, R.\/ de Ruyter van Steveninck and W.\/ Bialek.
\newblock {\em Spikes, exploring the neural code}.
\newblock The MIT Press, 1997.

\bibitem{schu-grass:96}
T.\/ Sch{\"{u}}rmann and P.\/Grassberger.
\newblock Entropy estimation of symbol sequences.
\newblock {\em Chaos}, 6:414--427, 1996.

\bibitem{Stevens:96}
C.F.\/ Stevens and A.\/Zador.
\newblock Information through a spiking neuron.
\newblock {\em NIPS}, 8, 1996.

\bibitem{Strong:98}
S.P.\/ Strong, R.\/ Koberle, R.\/ de Ruyter van Steveninck and W.\/ Bialek.
\newblock Entropy and information in neural spike trains.
\newblock {\em Physical Review Letters}, 80:197--200, 1998.

\bibitem{maxtree1}
P.A.J.\/ Volf and F.M.J.\/ Willems.
\newblock On the context tree maximizing algorithm.
\newblock {\em Proceedings of the 1995 IEEE International Symposium on 
  Information Theory}, 1995.

% \bibitem{statextbook}
% D.\/ Wackerly, W.\/ Mendenhall and R.\/ Scheaffer.
% \newblock {\em Mathematical statistics with applications}.
% \newblock Duxbury Press, 1996.

\bibitem{warland:97}
D.K.\/ Warland, P.\/ Reinagel and M.\/ Meister.
\newblock Decoding visual infomation from a population of retinal
ganglion cells.
\newblock {\em J. of Neurophysiology}, 78(5):2336--2350, 1997.

\bibitem{willems:95}
F.M.J.\/ Willems, Y.M.\/ Shtarkov and T.J.\/ Tjalkens.
\newblock The Context-tree weighting method: basic properties.
\newblock {\em IEEE Trans. Inform. Theory}, 41:653--664, 1995.

% \bibitem{willems:96}
% F.M.J.\/ Willems, Y.M.\/ Shtarkov and T.J.\/ Tjalkens.
% \newblock Context weighting for general finite-context sources.
% \newblock {\em IEEE Trans. Inform. Thoery}, 42:1514--1520, 1996.

\bibitem{willems:98}
F.M.J.\/ Willems.
\newblock The Context-tree weighting method: extensions.
\newblock {\em IEEE Trans. Inform. Theory}, 44:792--798, 1998.

\bibitem{ziv-lempel:1}
J.\/ Ziv and A.\/ Lempel.
\newblock A universal algorithm for sequential data compression.
\newblock {\em IEEE Trans. Inform. Theory}, 23:337--343, 1977.

% \bibitem{Ziv:78}
% J.\/ Ziv and A.\/ Lempel.
% \newblock Compression of individual sequences via variable rate coding.
% \newblock {\em IEEE Trans. Inform. Theory}, 24:530--536, 1978.

\end{thebibliography}
